\documentstyle[12pt,epsfig]{article}
\begin{document}
\renewcommand{\theequation}{\thesection.\arabic{equation}}
\newcommand{\eqn}[1]{(\ref{#1})}
\renewcommand{\section}[1]{\addtocounter{section}{1}
\vspace{5mm} \par \noindent
  {\large \bf \thesection . #1}\setcounter{subsection}{0}
  \par
   \vspace{2mm} } 
\newcommand{\sectionsub}[1]{\addtocounter{section}{1}
\vspace{5mm} \par \noindent
  {\bf \thesection . #1}\setcounter{subsection}{0}\par}
\renewcommand{\subsection}[1]{\addtocounter{subsection}{1}
\vspace{2.5mm}\par\noindent {\bf  \thesubsection . #1}\par
 \vspace{0.5mm} }
\renewcommand{\thebibliography}[1]{ {\vspace{5mm}\par \noindent{\bf
References}\par \vspace{2mm}}
\list
 {\arabic{enumi}.}{\settowidth\labelwidth{[#1]}\leftmargin\labelwidth
 \advance\leftmargin\labelsep\addtolength{\topsep}{-4em}
 \usecounter{enumi}}
 \def\newblock{\hskip .11em plus .33em minus .07em}
 \sloppy\clubpenalty4000\widowpenalty4000
 \sfcode`\.=1000\relax \setlength{\itemsep}{-0.4em}}
\def\a{& \hspace{-5pt}}
\def\dslash{\raisebox{1pt}{$\slash$} \hspace{-7pt} \partial}
\def\Dslash{\raisebox{1pt}{$\slash$} \hspace{-8pt} D}
\def\dslashs{\raisebox{0.75pt}{$\scriptstyle{\slash}$} \hspace{-4.5pt} \partial}
\def\Dslashs{\raisebox{0.75pt}{$\scriptstyle{\slash}$} \hspace{-5.5pt} D}
\newcommand\ubar[1]{#1 \hspace{-5.5pt} \raisebox{-2.5pt}{$\scriptscriptstyle{-}$}}
\newcommand\uubar[1]{#1 \hspace{-8pt} \raisebox{-2.75pt}{$\scriptstyle{-}$}}
\newcommand\uuubar[1]{#1 \hspace{-10pt} \raisebox{-3pt}{$\scriptstyle{-}$}
\hspace{-4pt} \raisebox{-3pt}{$\scriptstyle{-}$ \vspace{10pt}}}
\newcommand\uuuubar[1]{#1 \hspace{-15pt} \raisebox{-5pt}{$\scriptstyle{-}$}
\hspace{-4pt} \raisebox{-3pt}{$\scriptstyle{-}$ \vspace{10pt}}}
\def\bea{\begin{eqnarray}}
\def\eea{\end{eqnarray}}
\def\be{\begin{equation}}
\def\ee{\end{equation}}
\def\alp{\alpha}
\def\bet{\beta}
\def\gam{\gamma}
\def\del{\delta}
\def\eps{\epsilon}
\def\sig{\sigma}
\def\lam{\lambda}
\def\Lam{\Lambda}
\def\m{\mu}
\def\n{\nu}
\def\r{\rho}
\def\s{\sigma}
\def\d{\delta}
\newcommand\rf[1]{(\ref{#1})}
\def\nn{\nonumber}
\newcommand{\sect}[1]{\setcounter{equation}{0} \section{#1}}
\renewcommand{\theequation}{\thesection .\arabic{equation}}
\newcommand{\NPB}[3]{{Nucl.\ Phys.} {\bf B#1} (#2) #3}
\newcommand{\CMP}[3]{{Commun.\ Math.\ Phys.} {\bf #1} (#2) #3}
\newcommand{\PRD}[3]{{Phys.\ Rev.} {\bf D#1} (#2) #3}
\newcommand{\PLB}[3]{{Phys.\ Lett.} {\bf B#1} (#2) #3}
\newcommand{\JHEP}[3]{{JHEP} {\bf #1} (#2) #3}
\newcommand{\ft}[2]{{\textstyle\frac{#1}{#2}}\,}
\def\e{\epsilon}
\def\st{\scriptstyle}
\def\mco{\multicolumn}
\def\epp{\epsilon^{\prime}}
\def\vep{\varepsilon}
\def\ra{\rightarrow}
\def\ab{\bar{\alpha}}
\newcommand{\dt}{\partial_{\langle T\rangle}}
\newcommand{\dtbar}{\partial_{\langle\bar{T}\rangle}}
\newcommand{\al}{\alpha^{\prime}}
\newcommand{\mst}{M_{\scriptscriptstyle \!S}}
\newcommand{\mpl}{M_{\scriptscriptstyle \!P}}
\newcommand{\dv}{\int{\rm d}^4x\sqrt{g}}
\newcommand{\lv}{\left\langle}
\newcommand{\rv}{\right\rangle}
\newcommand{\ph}{\varphi}
\newcommand{\sbar}{\,\bar{\! S}}
\newcommand{\xbar}{\,\bar{\! X}}
\newcommand{\fbar}{\,\bar{\! F}}
\newcommand{\zbar}{\,\bar{\! Z}}
\newcommand{\tbar}{\bar{T}}
\newcommand{\ybar}{\bar{Y}}
\newcommand{\phb}{\bar{\varphi}}
\newcommand{\cm}{Commun.\ Math.\ Phys.~}
\newcommand{\pr}{Phys.\ Rev.\ D~}
\newcommand{\prl}{Phys.\ Rev.\ Lett.~}
\newcommand{\pl}{Phys.\ Lett.\ B~}
\newcommand{\ibar}{\bar{\imath}}
\newcommand{\jbar}{\bar{\jmath}}
\newcommand{\np}{Nucl.\ Phys.\ B~}
\newcommand{\gsi}{\,\raisebox{-0.13cm}{$\stackrel{\textstyle
>}{\textstyle\sim}$}\,}
\newcommand{\lsi}{\,\raisebox{-0.13cm}{$\stackrel{\textstyle
<}{\textstyle\sim}$}\,}

\thispagestyle{empty}

\begin{center}
\hfill SPIN-1999/05\\
\hfill LMU-TPW 99-07\\[3mm]
\hfill{\tt hep-th/9903145}\\

\vspace{2cm}

{\Large\bf Anomalies and inflow on D-branes and O-planes}\\[3mm]
\vspace{1.4cm}
{\sc Claudio A. Scrucca$^{a}$ and Marco Serone$^b$} \\
\vspace{1.3cm}

${}^a$
{\em Department of Physics, Ludwig Maximilian University of Munich}\\
{\em Theresienstra\ss e 37, 80333 Munich, Germany}\\
{\footnotesize \tt Claudio.Scrucca@physik.uni-muenchen.de}

\vspace{.5cm}

${}^b$
{\em Department of Mathematics, University of Amsterdam}\\
{\em Plantage Muidergracht 24, 1018 TV Amsterdam} \\
\& \\ {\em Spinoza Institute, University of Utrecht} \\
{\em Leuvenlaan 4, 3584 CE Utrecht, The Netherlands} \\
{\footnotesize\tt serone@wins.uva.nl}\\

\end{center}

\vspace{0.8cm}

\centerline{\bf Abstract}
\vspace{2 mm}
\begin{quote}\small
We derive the general form of the anomaly for chiral spinors and self-dual 
antisymmetric tensors living on D-brane and O-plane intersections,
using both path-integral and index theorem methods.
We then show that the anomalous couplings to RR forms of 
D-branes and O-planes in a general background are precisely those required to cancel 
these anomalies through the inflow mechanism. This allows, for instance, 
for local anomaly cancellation in generic orientifold models, the relevant Green-Schwarz term 
being given by the sum of the anomalous couplings of all the D-branes and 
O-planes in the model.  
\end{quote}
\newpage
\setcounter{equation}{0}


\section{Introduction}

One of the most important discoveries in the last few years of intense developments in string
theory is that Dp-branes and Op-planes carry the elementary RR p-form charges $\mu_p$ and 
$\mu_p^\prime = -2^{p-4} \mu_p$ \cite{pol}. It has also become clear that in a generic string 
background new charges with respect to lower RR forms are induced. For example, a
topologically non-trivial gauge bundle induces charges with respect to (p-2n)-forms \cite{li,dou}, 
whereas the curvature of the tangent and normal bundles induces charges with respect to 
(p-4n)-forms \cite{bsv,ghm,cy,mm,djm}. All these induced couplings are anomalous with respect 
to gauge transformations of the background, and are expected to cancel possible anomalies on the 
defects through the inflow mechanism \cite{ch}. This fact was indeed exploited to 
derive the complete anomalous couplings of a Dp-brane by requiring them to cancel the anomaly 
arising in the generically chiral world-volume theory on the intersection of two or more D-branes 
\cite{ghm,cy}. The presence of anomalous couplings for O-planes was instead predicted 
by string dualities \cite{djm}, whereas their relevance for anomaly cancellation has been argued in 
particular situations \cite{dm}. 

In \cite{mss}, a direct string computation of the complete anomalous 
couplings for Dp-branes and Op-planes has been given by factorizing magnetic interactions in 
a generic string background, confirming the indirect predictions for Dp-branes and correcting 
and extending those for Op-planes. The results are
\bea
S_{D_p} \a=\a \frac {\mu_{p}}{2} \int \,C\wedge\, \mbox{ch}_{\lambda} (F)\, \wedge \, \left.
\sqrt{\frac{\widehat{A}(R)}{\widehat{A}(R^\prime)}}\right|_{(p+1)-form}
\label{WZD} \\
S_{O_p}\a=\a \frac{\mu^\prime_p}{2} \int \,C \wedge \, \left.
\sqrt{\frac{\widehat{L}(R/4)}{\widehat{L}(R^\prime/4)}} \right|_{(p+1)-form} 
\label{WZO}
\eea
and have been further checked through disk computations \cite{cr,stef}, and indirectly
in other contexts (see for instance \cite{APTY}).
In these formulae, $C=\oplus_n\, C_{(n)}$ is the sum over the pulled-back RR form potentials.
$F$ is the field strength of the D-brane gauge field in the Chan-Paton 
factors representation $\lambda$ and $R,R^\prime$ are the pulled-back curvature 
two-forms of the tangent and normal bundles to the world-volume, all in 
units of $4 \pi^2 \al$. The overall $1/2$ normalization has been chosen in order to use later on
standard results for the inflow of anomaly, along the lines of \cite{cy}. The symbol 
$\mbox{ch}_\rho$ indicates the Chern class in the representation $\rho$,
\be
\mbox{ch}_\rho (F) = \mbox{Tr}_\rho \exp i \frac F {2 \pi} \;,
\ee
and $\widehat A(R)$ and $\widehat L(R)$ are the Roof genus and Hirzebruch polynomials,
given by
\be
\widehat A(R) = \prod_i \frac {\lambda_i/4 \pi}
{\sinh \lambda_i/4 \pi} \;,\;\;
\widehat L(R) = \prod_i \frac {\lambda_i/2 \pi}
{\tanh \lambda_i/2 \pi} \;,
\ee
in terms of the skew-eigenvalues $\lambda_i$ of $R_{\mu\nu}$. In considering the inflow 
mechanism, also the Euler class will appear:
\be
e(R) = \prod_i \lambda_i/2 \pi \;.
\ee

The aim of this paper is to show that the couplings (\ref{WZD}) and (\ref{WZO}), 
and similar additional model-dependent anomalous couplings to be determined case by case,
lead to an anomaly inflow on D-brane and O-plane intersections which precisely cancel
the anomalies on the corresponding world-volumes. 
From a string theory point of view, the only diagrams which can potentially give an anomaly 
are the divergent ones. At the one-loop level, these are the annulus, M\"obius strip and 
Klein bottle, associated respectively to brane-brane (BB), brane-orientifold (BO) and 
orientifold-orientifold (OO) intersections. The torus is instead always finite, and therefore can 
not give any anomaly. This means that even the anomaly of closed string fields living in the 
bulk of spacetime is located on orientifold planes. Quite in general, it is easy to figure out what kind of 
anomalous fields can appear on BB, BO and OO intersections. 
The cylinder and the M\"obius strip are surfaces with boundaries, corresponding to loops of 
arbitrary bosonic and fermionic open string states in the Ramond (R) and Neveu-Schwarz
(NS) sectors; the only massless 
anomalous particle which can arise on BB and BO intersections is therefore a charged chiral
R spinor reduced form $D=10$ to $d=p+1$. The Klein bottle is instead a surface without any 
boundary, and corresponds to loops of arbitrary bosonic closed string states in the RR and 
NSNS sectors; the only massless anomalous particle which can arise on OO intersections is 
therefore a neutral RR self-dual antisymmetric tensor, again reduced from $D=10$ to $d=p+1$. 

A very interesting consequence of this observation is that the anomalies arising from neutral 
closed string states have to combine to reproduce that of a self-dual tensor, whereas the anomalies
coming from charged open string states that of chiral spinors in suitable representations of the 
gauge group, as we will see. A slight clarification is here needed. In string theory, physical states 
generically arise after suitable truncations of the spectrum, implemented by projections 
like GSO, $\Omega$ or orbifold projections. Correspondingly, physical $g$-loop amplitudes 
are obtained by summing the contributions of all possible spin-structures on genus $g$ sufaces.
In particular, anomalies appear in the parity-violating part of one-loop amplitudes.
In string theory, this means genus one surfaces in the odd spin-structures, and restricting
the analysis to these spin-structures one can eventually identify which states,
among those propagating in the loop, are responsible for potential anomalies. 
As we have argued, these are a chiral spinor on BB and BO intersections and a self-dual tensor 
on OO intersections. The point is that these states {\it do not need} to appear in the physical 
string spectrum; they might be projected out in the truncated theory, but neverthless appear in 
different surfaces contributing to the same amplitudes. On the other hand, the inflow mechanism 
which will cancel these anomalies, that is nothing but the tree-level closed channel interpretation 
of the same diagrams, has to involve the exchange of physical RR forms only, appearing indeed in 
the effective action.

As a concrete example, consider for instance Type IIB orientifolds. Since the Type IIB theory 
one starts with in $D=10$ is chiral but anomaly free, any orientifold construction satisfying all 
consistency requirements of string theory, like in particular tadpole cancellation,
has to yield a theory which is automatically free of anomalies. However, it is 
well known that the massless fields arising in a generic orientifold model will in general give a 
non-vanishing anomaly. This implies that the D-branes and O-planes present in the
model have to contribute an equal and opposite anomaly through the inflow mechanism. 
This is nothing but a generalization of the Green-Schwarz (GS) mechanism \cite{GS}, the GS term 
being given by the sum of the anomalous couplings of all the D-branes and O-planes in the model, 
including possible additional anomalous couplings to RR forms coming from twisted string sectors.

Having identified the potentially anomalous states living on D-brane/O-plane intersections,
we will consider in a more general context the possible anomalies arising from chiral spinors
and self-dual antisymmetric tensors propagating in a given submanifold of spacetime, but in
interaction with gravity of all spacetime.

The plan of the paper is the following. In Section 2 we compute the general form of the anomaly 
for chiral spinors and self-dual antisymmetric tensors coupling to both the tangent and the normal 
bundle curvatures, generalizing some standard results \cite{agw} (see also \cite{agg} for an
extended review of anomalies in field theory). Using Fujikawa's approach 
\cite{fuj}, we first relate these anomalies to chiral anomalies, which can then be interpreted 
as indices of classical complexes endowed with an additional group action. 
We then compute these indices in 
two different ways. The first more physical approach consists in regularizing the index as the 
high-temperature limit of the partition function of a suitable supersymmetric theory, as in 
\cite{ag,agw,fw}. The second is more mathematical and relies on the application of the 
so-called index and G-index theorems \cite{AS,ASg}.
Although the general form of the anomaly for chiral spinors has been already obtained in \cite{cy}
by using the family index theorem \cite{ASf}, we will re-derive and 
confirm this result in the next sections 
in the two different ways mentioned above. In particular, the explicit computation based on a 
supersymmetric quantum mechanics, obtained as the reduction of the supersymmetric non-linear 
$\sigma$-model \cite{sigma} from 1+1 to 0+1 dimensions with some constraints 
on the fields, will turn out to be very instructive due to its close connection to the 
open string $\sigma$-model in a  curved D-brane background. In Section 3 we give a detailed 
description of the anomaly inflow on D-brane/O-plane intersections, showing that it always 
precisely cancels world-volume field theory anomalies. In Section 4 we
apply our results to some Type IIB orientifold models, discuss general features of 
anomaly cancellation in these models and finally in last section we give some conclusions.


\section{Anomalies for spinors and tensors}
\setcounter{equation}{0}

As explained in the introduction, we need to compute the anomaly for a chiral spinor field and
a self-dual antisymmetric tensor field (that is with a  self-dual field strength) reduced from some 
$D$-dimensional manifold $X$ to a submanifold $M\subset X$ of dimension 
$d<D$. More precisely, with the term reduced we mean here the generalization to a 
non-trivial normal bundle of the usual dimensional reduction. 
Recall that upon dimensional reduction from $D$ to $d$ dimensions, the $D$-dimensional 
Lorentz group is broken to the d-dimensional Lorentz group plus an R-symmetry corresponding 
to rotations in the $(D-d)$-dimensional transverse space: $SO(D-1,1) \rightarrow SO(d-1,1) 
\times SO(D-d)$. The tangent bundle to $X$ restricted to $M$ decomposes as the 
Whitney sum of the tangent and normal bundles to $M$: 
\be
T(X)|_M = T(M) \oplus N(M) \;.
\label{whit}
\ee 
Correspondingly, a field in $D$ dimensions in some representation $R$ of
$SO(D-1,1)$ decomposes into various multiplets of fields in $d$ dimensions, in representations 
$(R_1^i,R_2^i)$ of $SO(d-1,1) \times SO(D-d)$. More precisely, a section of $T(X)$ in some 
representation $R$ will decompose into sections of $T(M) \otimes N(M)$ in representations 
$(R_1^i,R_2^i)$. For simplicity and motivated by D-brane physics, we will consider fields 
that couple only to gravity on $X$, restricting any possible gauge field to connections
on bundles over $M$.

Under the above reduction, local Lorentz symmetry (or general covariance) on $X$ is broken
to local Lorentz symmetry on $M$ and local R-symmetry on the tranverse space. The former is just 
standard gravity seen as a gauge theory whose gauge field is the spin connection on the tangent 
bundle $T(M)$, the fields transforming in the $R_1^i$ representations. The latter corresponds instead 
to a gauge symmetry under which the fields transform in the $R_2^i$ representations, the
gauge field associated to this gauge symmetry being simply the spin connection on the normal 
bundle $N(M)$. Any anomalous representation $R$ will decompose into pairs 
of representations $(R_1^i,R_2^i)$ and $(\bar R_1^i, \bar{R}_2^i)$ related by conjugation.
If $N(M)$ is trivial, the representations $R_2^i$ and $\bar{R}_2^i$ are equivalent and the two 
components $R_1^i$ and $\bar R_1^i$ will give an equal and opposite contribution to the anomaly, 
which will therefore vanish. On the other hand, if $N(M)$ is non-trivial, so are the bundles lifted 
from it in the $R_2^i,\bar{R}_2^i$ representations, and the latter are no longer 
equivalent; the two components $R_1^i$ and $\bar R_1^i$ can then give unbalanced contributions
and the anomaly can be non-vanishing. 

In the following we use the standard notation for Wess-Zumino descents. 
The anomaly is encoded in a closed and gauge-invariant sum of forms $I$, function of the 
curvature 2-forms of the gauge, tangent and normal bundles. Apart from a 
possible constant term $I_0$,  this will also be exact, since so are the curvatures:  $I-I_0 = dI^{(0)}$. 
$I^{(0)}$ is not gauge invariant; rather its gauge variation defines the Wess-Zumino
descent $I^{(1)}$: $\delta_\eta I^{(0)} = dI^{(1)}$. Anomalies in field theory always have the form 
$A= 2 \pi i \int_M I^{(1)}$; this ensures that the Wess-Zumino consistency condition is automatically 
satisfied. 

\subsection{Path-integral computation}

\vskip 2pt

As anticipated, we will first use Fujikawa's method \cite{fuj} to compute the anomaly for a reduced 
chiral spinor and self-dual antisymmetric tensor. In this approach, the anomaly is attributed to the 
Jacobian arising from the non-invariant path-integral measure under a generic background gauge 
transformation $\delta_\eta$. This method presents the advantage of being very easily generalized 
to the case in which the field is reduced to a lower dimensional manifold. 
It will prove very convenient to use the strategy of looking at the reduced
case as the unreduced case with a constraint. Correspondingly, the generic gauge variation 
$\delta_\eta$ contains gauge transformations and reparametrizations of both the tangent and the 
normal bundle. Since the dependence of the anomaly on the gauge field in the reduced case is the
same as in the standard unreduced case, we shall concentrate in the following on the gravitational 
part.

In the spirit of \cite{agw}, we regularize the ill-defined traces encoding the anomaly 
as the high-temperature limits of partition functions of suitable supersymmetric theories. 
These will turn out to be different versions of the supersymmetric non-linear 
$\sigma$-model \cite{sigma} reduced from 1+1 down to 0+1 dimensions, whose Lagrangian is
\be
L=\frac12 \left[g_{MN}(x)\dot{x}^M\dot{x}^N
+i\psi_{1\uubar{M}}D_{\tau}(x)\psi_1^{\uubar{M}}+i\psi_{2\uubar{M}}D_{\tau}(x)\psi_2^{\uubar{M}}
+\frac12 R_{\uubar M\uubar N\uubar P\uubar Q}(x)
\psi_1^{\uubar{M}}\psi_1^{\uubar{N}}\psi_2^{\uubar{P}}\psi_2^{\uubar{Q}}
\right] \label{S0}
\ee
where
\be
D_{\tau}\psi_{\alpha}^{\uubar{M}}=\dot \psi_{\alpha}^{\uubar{M}}+
\omega_{M\;\;\uubar{N}}^{\;\;\;\uubar{M}}(x)\,\psi_{\alpha}^{\uubar{N}}\,\dot{x}^M
\;,\;\; \alpha=1,2 \;.
\ee
In the action (\ref{S0}), we have introduced the Lorentz frame fermion fields 
$\psi_{\alpha}^{\uubar{M}}=e^{\uubar{M}}_{\;\;M}\,\psi_{\alpha}^M$. Here and in the 
following capital indices $M,N,...$ run over the total space-time, whereas Greek and Latin indices
$\mu,\nu,...$, $i,j,...$ denote respectively coordinates on $M$ and the transverse space to it in $X$; 
underlined indices denote instead flat indices, non-underlined ones being curved.

\subsubsection{Chiral spinor}

Consider first a chiral spinor. 
In the usual unreduced case, the Jacobian giving the gauge and/or the gravitational anomaly can be 
written as \cite{agw} 
\be
A_X = \lim_{t \rightarrow 0} \mbox{Tr} \,[\Gamma^{D+1} \delta_\eta 
e^{- t (i \Dslashs_X)^2}]
\label{Ax}
\ee
where $\Gamma^{D+1}$ is the chiral matrix in $D$ dimensions, the operator $\delta_{\eta}$ 
represents the corresponding gauge and/or gravitational variation  and the trace runs over the 
eigenstates of $H=(i \Dslash_X)^2$, $\Dslash_X$ being the Dirac operator on $X$. 
The operator $\delta_\eta$ can be exponentiated, resulting in a shift in the background 
fields. The final result has then to be restricted to the term which is linear in $\eta$. 
This procedure corresponds to the well known fact that gauge and gravitational 
anomalies can be obtained from chiral anomalies by taking the Wess-Zumino descent. 

In the reduced case, in which the chiral spinor propagates on $M$ but couples
through a gauge-like coupling to the normal bundle curvature, the anomaly is still given 
by (\ref{Ax}) but with $\Dslash_X$ replaced by the Dirac operator $\Dslash_M$ on $M$,
that now includes the connection on the normal bundle of $M$. Again, by exponentiating
the operator $\delta_\eta$, we are left with
\be
Z_M = \lim_{t \rightarrow 0} \mbox{Tr} \,[\Gamma^{D+1} e^{- t (i \Dslashs_M)^2}] \;.
\label{anospinor}
\ee
A useful way to evaluate (\ref{anospinor}) is by looking for a supersymmetric quantum mechanical
theory whose supercharge is $Q=i \Dslash_M$. 
It is not difficult to check that such a theory can
be obtained by reducing the supersymmetric non-linear $\sigma$-model 
from $1+1$ to $0+1$ dimensions, with Neumann boundary conditions for the fields in the
directions 
inside $M$ and Dirichlet boundary conditions for the fields in the remaining 
directions. 
In terms of the fields appearing in (\ref{S0}) this means
$x^i=0$, $\psi_1^{\ubar{\mu}} = \psi_2^{\ubar{\mu}}\equiv \psi^{\ubar{\mu}}/\sqrt{2}$ and 
$\psi_1^{\ubar{i}} = -\psi_2^{\ubar{i}}\equiv \psi^{\ubar{i}}/\sqrt{2}$.
The action (\ref{S0}) then leads to the following Lagrangian
\bea
L \a=\a \frac 12 g_{\mu\nu} \dot x^\mu \dot x^\nu 
+ \frac i2 \psi_{\ubar \mu} \left(\dot \psi^{\ubar \mu} 
+ \frac 12 \omega_{\rho \;\;\; \ubar \nu}^{\;\; \ubar \mu} \,\dot x^\rho\, \psi^{\ubar \nu} \right)
+ \frac i2 \psi_{\ubar i} \left(\dot \psi^{\ubar i}
+ \frac 12 \omega_{\rho \;\;\; \ubar j}^{\;\; \ubar i}\, \dot x^\rho\, \psi^{\ubar j} \right) \nn \\
\a\;\a + \frac 14 R_{\ubar \mu \ubar \nu \ubar i \ubar j} \psi^{\ubar \mu} \psi^{\ubar \nu} 
\psi^{\ubar i} \psi^{\ubar j} \;.
\label{spinormodel2}
\eea
This is still invariant under a combination of the two supersymmetries 
of (\ref{S0}). More precisely, the operator 
$Q= e^\mu_{\;\;\ubar\mu}\,\psi^{\ubar\mu}\,\dot{x}_\mu$ is still a conserved supercharge.

After canonical quantization,  the $\psi^{\ubar \mu}$'s and 
$\psi^{\ubar i}$'s satisfy the anticommutation relations 
$\{\psi^{\ubar \mu},\psi^{\ubar \nu}\} = \delta^{\ubar \mu \ubar \nu}$, 
$\{\psi^{\ubar i},\psi^{\ubar j}\} = \delta^{\ubar i \ubar j}$ and 
$\{\psi^{\ubar \mu},\psi^{\ubar i}\} = 0$,
and generate Clifford algebras on $M$ and the transverse space to it in $X$. They form 
therefore bases of forms respectively on $M$ and its transverse space.
The canonical momentum $\pi_\mu$ conjugate to $x^\mu$ is found to be
\be
\pi_\mu = g_{\mu\nu} \dot x^\nu + \frac i4 \left(\omega_{\mu}^{\;\; \ubar \nu \ubar \rho} 
[\psi_{\ubar \nu}, \psi_{\ubar \rho}] + \omega_{\mu}^{\;\; \ubar i \ubar j} 
[\psi_{\ubar i}, \psi_{\ubar j}] \right) \;.
\ee
Upon canonical quantization, $\pi_\mu \rightarrow -i \partial_\mu$ and 
$\psi_{\uubar M} \rightarrow \Gamma_{\uubar M}/\sqrt{2}$, and the supercharge 
becomes 
\be
Q = - \frac i{\sqrt{2}} e^\mu_{\;\ubar \mu} \Gamma^{\ubar \mu} \left[\partial_\mu + 
\frac 14 \left(\omega_{\mu}^{\;\; \ubar \nu \ubar \rho} \Gamma_{\ubar \nu \ubar \rho} + 
\omega_{\mu}^{\;\; \ubar i \ubar j} \Gamma_{\ubar i \ubar j} \right) \right]
\ee
which is indeed the Dirac operator on $M$: $Q = - \Dslash_M/\sqrt{2}$. The Lagrangian 
(\ref{spinormodel2}) is actually a particular case of the one which was found in \cite{ag}
to have as supercharge the Dirac operator with an arbitrary gauge connection.
Finally, the chiral matrix $\Gamma^{D+1}$, which can be interpreted also as 
$\Gamma^{d+1} \Gamma^{D-d+1}$, is represented by the fermion number operator $(-1)^F$.

The partition function (\ref{anospinor}) giving the anomaly is recognized to be the Witten 
index \cite{wit} for the theory described by the Lagrangian (\ref{spinormodel2}):
\be
\label{indexspinor}
Z_M = \mbox{Tr} \,[(-1)^F e^{- t H}] \;.
\ee
Actually, being a topological quantity, this index does not depend on $t$. Its functional 
integral representation is
\be
Z_M = \int_P \! {\cal D}x^\mu(\tau)  \int_P \! {\cal D} \psi^{\ubar\mu} (\tau) 
\int_P {\cal D} \! \psi^{\ubar i}(\tau) 
\exp \left\{- \int_0^t \! d\tau L \left(q^\mu(\tau),\psi^{\ubar \mu} (\tau),\psi^{\ubar i}(\tau) \right) \right\} \;.
\label{pathspinor}
\ee
Due to the $(-1)^F$ insertion, all the fields are periodic. In order to evaluate this path-integral, 
it will be convenient to take the high-temperature limit $t \rightarrow 0$. In that limit, 
(\ref{pathspinor}) is dominated by constant paths $x_0^\mu$,  $\psi_0^{\ubar \mu}$,
$\psi_0^{\ubar i}$, with minimal energy $E_0 = - R_{\ubar \mu \ubar \nu \ubar i \ubar j} (x_0) 
\psi_0^{\ubar \mu} \psi_0^{\ubar \nu} \psi_0^{\ubar i} \psi_0^{\ubar j}/4$
and Euclidean action $S_0 = t E_0$. The functional integral is evaluated by expanding the fields 
as fluctuations around these constant paths, $x^\mu = x_0^\mu + \xi^{\mu}$, 
$\psi^{\ubar \mu} = \psi_0^{\ubar \mu} + \lambda^{\ubar \mu}$ and 
$\psi^{\ubar i} = \psi_0^{\ubar i} + \lambda^{\ubar i}$. 
In the present case, as we will see,
it will be enough to consider quadratic fluctuations and perform a one-loop computation, since 
higher loop corrections come with additional powers of $t$ and are irrelevant in the limit 
$t \rightarrow 0$ we are considering. Expanding then the Lagrangian (\ref{spinormodel2}) in 
normal coordinates \cite{agfm} around $x_0^\mu$ and keeping only terms at most bilinear in the
fluctuations, one finds a quadratic action for the fluctuations which depends on the 
fermionic zero modes $\psi_0^{\ubar \mu}$ and $\psi_0^{\ubar i}$.
The path integral then reduces to the integral over the bosonic and fermionic zero modes 
of the determinants arising from the Gaussian integration over the fluctuations. The integral over the 
$x_0^\mu$'s is just the integral over the manifold $M$, whereas the integrals over the 
$\psi_0^{\ubar \mu}$'s and the $\psi_0^{\ubar i}$'s select the d-form component 
$\psi_0^{\ubar \mu_{\raisebox{2pt}{$\scriptscriptstyle{0}$}}} ... 
\psi_0^{\ubar \mu_{\raisebox{2pt}{$\scriptscriptstyle{d}$}}}$ on $M$ and 
the (D-d)-form component  $\psi_0^{\ubar i_{\raisebox{2pt}{$\scriptscriptstyle{d+1}$}}} ... 
\psi_0^{\ubar i_{\raisebox{2pt}{$\scriptscriptstyle{D-d}$}}}$ on the transverse space in $X$.

The quadratic Lagrangian for the fluctuations contains terms with two or less fermionic zero modes.
It is clear that, due to the integrals over fermionic zero modes, only interactions providing
the maximal number of them (that is 2) will be relevant; indeed, picking up other
interactions would increase the total number of vertices required to provide a sufficient
number of fermionic zero modes in order to get a non-vanishing result.
Among these, we now argue that the $t$-independent
term of the path integral depends (besides the constant term) only on the terms 
quadratic in the fluctuations and bilinear in the $\psi_0^{\ubar\mu}$'s.
The reason is the following:
the tree level term $L_0$ contains 4 zero modes and a single power of $t$, 
whereas quadratic vertices are also effectively proportional to $t$, as we will see,
but provide only 2 fermionic zero modes. The leading 
contribution to the path-integral for $t \rightarrow 0$ comes therefore from correlations 
involving a maximum number of tree-level terms.
If $d>D/2$, this term saturates alone the integral over the $\psi_0^{\ubar i}$'s, contributing with 
a power $t^{(D-d)/2}$ and providing $D-d$  $\psi_0^{\ubar\mu}$'s. Since the $d$ periodic bosons
$\xi^{\ubar\mu}$ give a normalization factor of order $t^{-d/2}$, the total power of $t$
becomes $t^{(D-2d)/2}$. We still need to soak up the remaining $2d-D$  
$\psi_0^{\ubar\mu}$'s. It is then clear that only the terms bilinears in the $\psi_0^{\ubar\mu}$'s
contribute to the leading $t$-independent term. Any other contribution vanishes in the limit 
$t\rightarrow 0$. This shows that we do not need to consider higher-loop contributions and 
that we can safely neglect all terms proportional to $\psi_0^{\ubar i}$'s in the quadratic Lagrangian.
If $d<D/2$, a similar line of arguments shows that the anomaly vanishes.

The effective Lagrangian, quadratic in the fluctuations and in the $\psi_0^{\bar \mu}$'s, is 
found to be
\bea
L^{eff} \a=\a \frac 12 \left(\dot \xi_{\ubar \mu} \dot \xi^{\ubar \mu} 
+ i \lambda_{\ubar \mu} \dot \lambda^{\ubar \mu} + i \lambda_{\ubar i} \dot \lambda^{\ubar i} 
+ i R_{\ubar \mu \ubar \nu} \dot \xi^{\ubar \mu} \xi^{\ubar \nu} 
+ R^\prime_{\ubar i \ubar j} \lambda^{\ubar i} \lambda^{\ubar j} 
+ R^\prime_{\ubar i \ubar j} \psi_0^{\ubar i} \psi_0^{\ubar j} \right) 
\label{spinoreff}
\eea
where we have defined the tangent and normal bundle curvature 2-forms as
\be
R_{\ubar\mu \ubar\nu} = \frac 12 R_{\ubar \mu \ubar \nu \ubar \rho \ubar \sigma} (x_0) 
\psi_0^{\ubar\rho} \psi_0^{\ubar\sigma} \;,\;\;
R^\prime_{\ubar i\ubar j} = \frac 12 R_{\ubar i \ubar j \ubar \rho \ubar \sigma} (x_0) 
\psi_0^{\ubar \rho} \psi_0^{\ubar \sigma} \;.
\label{tncurv}
\ee
The integral over the constant tree-level part of the action gives
\be
\int \! d\psi_0^{\ubar i} \exp \left\{t \sum_{i=d/2}^{D/2} R^\prime_{\ubar i \ubar j} 
\psi_0^{\ubar i} \psi_0^{\ubar j} \right\} = \prod_{i=d/2}^{D/2} \lambda^\prime_i t \;.
\label{treelevel}
\ee
The evaluation of the one-loop determinants is straightforward. Using $\zeta$-function 
regularization to carefully normalize them, one finds
\bea
\a\a\mbox{det}^{-1}_P(\partial_\tau^2 \eta_{\ubar \mu \ubar \nu} +
i R_{\ubar \mu \ubar \nu} \partial_\tau) = 
(2 \pi t)^{-\frac d2} \prod_{i=1}^{d/2} \frac {\lambda_i t/2}{\sinh \lambda_i t/2} \;, \nn \\
\a\a\mbox{det}_P(i \partial_\tau \eta_{\ubar \mu \ubar \nu}) = 1 \;,\;\;
\mbox{det}_P(i \partial_\tau \eta_{\ubar i \ubar j} + R^\prime_{\ubar i \ubar j}) = 
\prod_{i=d/2}^{D/2} \frac {\sinh \lambda^\prime_i t/2}{\lambda^\prime_i t/2} \;.
\eea

Note at this point that the inclusion of an arbitrary gauge bundle presents no difficulties. 
The supersymmetric quantum mechanical model (\ref{spinormodel2}) has to be extended in such a way 
that its supercharge acquires an additional term involving the gauge connection, reproducing 
therefore the Dirac operator for a charged spinor. This modification is achieved exactly in the 
same way as in the standard case \cite{agw}, and results in the additional tree-level  factor 
\be
\mbox{Tr}_\rho \exp \left\{i t F\right\}
\ee
where $\mbox{Tr}_\rho$ indicates the trace over the gauge group in the representation $\rho$ in 
which the spinor transforms and
\be
F = \frac 12 F_{\ubar \mu \ubar \nu} (x_0) \psi_0^{\ubar \mu} \psi_0^{\ubar \nu}
\ee
is the curvature 2-form of the gauge bundle.

Taking into account also the effect of the gauge field, the result for the path-integral 
(\ref{pathspinor}) is then:
\bea
Z_M \a=\a \int \! dx_0^\mu \int \! d\psi_0^{\ubar \mu} \,
(2 \pi t)^{-\frac d2} \mbox{Tr}_\rho \exp \left\{i t F \right\} 
\prod_{i=1}^{d/2} \frac {\lambda_i t/2}{\sinh \lambda_i t/2}
\prod_{i=d/2}^{D/2} \frac {\sinh \lambda^\prime_i t/2}{\lambda^\prime_i t/2}
\prod_{i=d/2}^{D/2} \lambda^\prime_i t \nn \\
\a=\a \int \! dx_0^\mu \int \! d\psi_0^{\ubar \mu} \, \mbox{Tr}_\rho \exp \left\{i F/2\pi \right\}
\prod_{i=1}^{d/2} \frac {\lambda_i/4\pi}{\sinh \lambda_i/4\pi}
\prod_{i=d/2}^{D/2} \frac {\sinh \lambda^\prime_i/4\pi}{\lambda^\prime_i/4\pi}
\prod_{i=d/2}^{D/2} \lambda^\prime_i/2\pi \hspace{20pt}
\eea
where in the last step we have used the fact that only the d-form component of the integrand 
contributes. This result can be rewritten as
\be
Z_M = \int_M \mbox{ch}_\rho (F) \wedge \frac {\widehat A (R)}{\widehat A (R^\prime)} \wedge 
e(R^\prime) \;.
\label{Aspi}
\ee
This is the chiral anomaly of a spinor propagating on $M$ but section of the
spin bundle of $X\supset M$. As last step, we have to consider how the operator
$\delta_\eta$ is realized, in order to derive the form of gauge and gravitational anomalies
on $M$. Reparametrizations in $X$ are broken down to reparametrizations of $M$ and rotations
in the transverse space. The former correspond to tangent bundle gauge transformations, 
$\delta_\eta \psi_a = - \eta^\mu D_\mu \psi_a$, and the latter to normal bundle
gauge transformations, $\delta_{\eta^\prime} \psi_a = - D_{\ubar i}
\eta^\prime_{\ubar j} \Gamma^{\ubar i \ubar j}_{ab} \psi_b$.
It is not difficult to verify that the operators $\delta_\eta$ and 
$\delta_{\eta^\prime}$ are represented by
$$
\delta_\eta = - \eta_\mu \dot x^\mu \;,\;\; \delta_{\eta^\prime} = 
- D_{\ubar i} \eta^\prime_{\ubar j} \psi^{\ubar i} \psi^{\ubar j}
$$
after canonical quantization. It is then easy to show that exponentiating these operators and
expanding in normal coordinates, their net effect in (\ref{spinoreff}) is to shift 
$R_{\mu\nu} \rightarrow R_{\ubar \mu \ubar \nu} + D_{\ubar \mu}
\eta_{\ubar \nu} - D_{\ubar \nu} \eta_{\ubar \mu}$ and 
$R_{\ubar i \ubar j} \rightarrow R_{\ubar i \ubar j} + D_{\ubar i}
\eta_{\ubar j} - D_{\ubar j} \eta_{\ubar i}$.
Taking the terms linear in $\eta,\eta^\prime$ corresponds therefore to
take the Wess-Zumino descent, and the final result for the complex chiral spinor
anomaly turns out to be, as expected:
\be
A_M = 2 \pi i \int_M \left[ \mbox{ch}_\rho (F) \wedge \frac {\widehat A (R)}{\widehat A (R^\prime)} 
\wedge e(R^\prime) \right]^{(1)} \;.
\label{Aspi(1)}
\ee

\subsubsection{Self-dual tensor}

Consider now a self-dual tensor interacting with gravity on $X$. 
The Jacobian giving the anomaly has been shown to be given by \cite{agw}
\be
A_X = \lim_{t \rightarrow 0} \,\frac 14\,\mbox{Tr} \,[*_D \delta_\eta e^{- t \Box_X}]
\label{ten}
\ee
where $*_D$ is the Hodge operator in $D$ dimensions and the trace runs over the eigenstates of 
$H=\Box_X$, with $\Box_X = (d + d^\dagger)^2$ the Laplace-Beltrami operator on $X$. 
We want now to consider gravitational anomalies on $M\subset X$, arising by reducing
a chiral tensor from $D$ down to $d$ dimensions.
The corresponding expression is obtained from (\ref{ten}) by tracing only over the 
states propagating on $M$. 
However, it is not possible to follow the same strategy as in the case of the chiral spinor,
essentially because there is no simple theory having the required Hamiltonian. Indeed,
it is well known that the supersymmetric action (\ref{S0}) 
has a Hamiltonian which is the Laplacian on forms of the target space 
\cite{wit}, but there is no evident deformation of it which could constrain the dynamics to a 
submanifold of the target space. Fortunately, there is a second way of restricting the trace,
which will make possible the evaluation of the anomaly: rather than changing the Hamiltonian,
one can insert in the trace a suitable operator having the effect of projecting out the 
unwanted eigenstates. More precisely, we want to keep only those states with vanishing 
momentum in the transverse space to $M$ in $X$. The appropriate operator to insert turns 
out to be quite simple and given by the reflection operator $I$ in the transverse 
space. Since $I$ acts on the fields $x^M$ and $\psi^{\uubar M}$ 
in (\ref{S0}) with a $+$ and a $-$ sign in the tangent and normal directions to $M$, 
states with non vanishing momentum $p^i$
will be reflected into orthogonal states with opposite momentum $-p^i$, giving so a vanishing 
contribution to the trace.
We can therefore extend the trace on the whole set of states propagating on $X$
\be
A_M = \lim_{t \rightarrow 0} \mbox{Tr} \,[I *_D \delta_\eta e^{- t \Box_X}] \;.
\label{Am}
\ee
The trace (\ref{Am}) is actually taken, as in the standard case, over all the 
tensor fields (differential forms) on $X$. However, it is clear that the only non-vanishing
contribution to (\ref{Am}) comes from $d/2$-forms on $M$, arising from the
reduction of the self-dual $D/2$-form on $X$.

As before, the operator $\delta_\eta$ can be exponentiated, resulting in a shift in the background 
fields. We are then left with the ``self-duality anomaly''
\be
\label{anotensor}
Z_M = \lim_{t \rightarrow 0} \mbox{Tr} \,[I *_D e^{- t \Box_X}] \;.
\ee
Although the operator $I$ {\it does not} commute in general with the Hamiltonian $\Box_X$,
according to the decomposition (\ref{whit}), it commutes with the Laplace-Beltrami
operator restricted to $M$. Using standard arguments \cite{wit}, it is then 
clear
that the trace (\ref{anotensor}) will be again an index. 
In order to evaluate (\ref{anotensor}), we regard it as the partition function for 
the supersymmetric quantum mechanical model (\ref{S0}), whose Hamiltonian is $H=\Box_X$.
The $*_D$ operator, which can be interpreted now as $*_d *_{D-d}$, is implemented 
as usual by the discrete symmetry $\Omega$, mapping $\psi_1\rightarrow - \psi_1$
and $\psi_2\rightarrow \psi_2$. 
Only zero energy states can contribute to (\ref{anotensor}); indeed 
all the massive ones fall into multiplets
with equal number of eigenstates of $I \Omega$ with opposite eigenvalues, and the result is 
therefore independent of $t$. The path integral representation of (\ref{anotensor}) is 
\bea
Z_M \a=\a \int_P \! {\cal D}x^\mu(\tau) \int_A \! {\cal D}x^{i}(\tau)  
\int_P \! {\cal D} \psi_1^{\ubar\mu} (\tau)  \int_A \! {\cal D} \psi_1^{\ubar i} (\tau)
\int_A {\cal D} \! \psi_2^{\ubar \mu}(\tau) \int_P {\cal D} \! \psi_2^{\ubar i}(\tau) \nn \\
\a\;\a \exp \left\{- \int_0^t \! d\tau L \left(x^M(\tau),\psi_{1,2}^{\ubar\mu} (\tau),
\psi_{1,2}^{\ubar i}(\tau) \right) \right\}
\label{pathtensor}
\eea
where the Lagrangian is given by (\ref{S0}).
The periodicities are obtained by noting that $\Omega I$ acts with a $+$ sign on $x^\mu$,
$\psi_1^{\ubar i}$, $\psi_2^{\ubar \mu}$ and with a $-$ sign on $x^{ i}$, 
$\psi_1^{\ubar \mu}$, $\psi_2^{\ubar i}$. Notice
that only $x^\mu$,  $\psi_1^{\ubar \mu}$ and $\psi_2^{\ubar i}$ have then zero modes.
Again, it is convenient to take the high-temperature limit $t \rightarrow 0$. In that limit, 
(\ref{pathtensor}) is dominated by constant paths and one can therefore 
expand the fields as $x^\mu = x_0^\mu + \xi^{ \mu}$, 
$x^i = \xi^{i}$, $\psi_1^{\ubar \mu} = \psi_{0}^{\ubar \mu} + \lambda_1^{\ubar \mu}$, 
$\psi_2^{\ubar \mu} = \lambda_2^{\ubar \mu}$, 
$\psi_1^{\ubar i} =  \lambda_1^{\bar i}$ and 
$\psi_2^{\ubar i} = \psi_{0}^{\ubar i} + \lambda_2^{\ubar i}$.
Expanding the Lagrangian (\ref{S0}) in normal coordinates around $x_0^\mu$, 
it is evident that the last term in (\ref{S0}) will again give the same tree-level 
term as in the spinor case: $L_0 =  t R_{\ubar \mu \ubar \nu \ubar i \ubar j} (x_0) 
\psi_0^{\ubar \mu} \psi_0^{\ubar \nu}\psi_0^{\ubar i} \psi_0^{\ubar j}/4$. 
Then, by applying precisely the same considerations 
as in the spinor case, the $t$-independent term in the path integral (\ref{pathtensor}) will 
receive contributions only from the tree-level term and quadratic interactions bilinear
in the $\psi_0^{\ubar\mu}$'s. The effective Lagrangian that one obtains is then
\bea
L^{eff} \a=\a \frac 12 \left[\dot \xi_{\ubar \mu} \dot \xi^{\ubar \mu} + 
\dot \xi_{\ubar i} \dot \xi^{\ubar i}
+ i \lambda_{1\ubar \mu} \dot \lambda_1^{\ubar \mu}  + i \lambda_{1\ubar i} \dot \lambda_1^{\ubar i}
+ i \lambda_{2\ubar \mu} \dot \lambda_2^{\ubar \mu}  + i \lambda_{2\ubar i} 
\dot \lambda_2^{\ubar i} \right. \nn \\ 
\a\;\a \hspace{12pt} \left. + R_{\ubar \mu \ubar \nu} \left(i\,\dot \xi^{\ubar\mu} \xi^{\ubar \nu}
+ \lambda_2^{\ubar \mu}  \lambda_2^{\ubar \nu} \right) + 
R^\prime_{\ubar i \ubar j} \left(i\,\dot \xi^{\ubar i} \xi^{\ubar j} 
+ \lambda_2^{\ubar i}  \lambda_2^{\ubar j} \right)  
+ R^\prime_{\ubar i \ubar j} \psi_0^{\ubar i} \psi_0^{\ubar j} \right]
\label{tensoreff}
\eea
in terms of the tangent and normal bundle curvature 2-forms (\ref{tncurv}).

The constant tree-level part of the action 
again contributes as in (\ref{treelevel}). The evaluation of the one-loop 
determinants presents no difficulties, a part from important normalizations which can be
fixed by relying again on $\zeta$-function regularization. One finds in this way 
the following results for the two bosonic and four fermionic fluctuations:
\bea
\a\a\mbox{det}^{-1}_P(\partial_\tau^2 \eta_{\ubar \mu \ubar \nu} +i\,
R_{\ubar \mu \ubar \nu} \partial_\tau) = 
(2 \pi t)^{-\frac d2} \prod_{i=1}^{d/2} \frac {\lambda_i t/2}{\sinh \lambda_i t/2} \;, \nn \\
\a\a\mbox{det}^{-1}_A(\partial_\tau^2 \eta_{\bar i \bar j} + i\, R_{\bar i \bar j} \partial_\tau) = 
\prod_{i=d/2}^{D/2} \frac 1{4 \cosh \lambda^\prime_i t/2} \;, \nn \\
\a\a\mbox{det}_P(i \partial_\tau \eta_{\ubar \mu \ubar \nu}) = 1 \;,\;\;
\mbox{det}_A(i \partial_\tau \eta_{\ubar \mu \ubar \nu} + R_{\ubar \mu \ubar \nu}) = 
\prod_{i=1}^{d/2} 2 \cosh \lambda_i t/2 \;, \nn \\
\a\a\mbox{det}_A(i \partial_\tau \eta_{\ubar i \ubar j}) = \prod_{i=d/2}^{D/2} 2 \;,\;\;
\mbox{det}_P(i \partial_\tau \eta_{\ubar i \ubar j} + R^\prime_{\ubar i \ubar j}) = 
\prod_{i=d/2}^{D/2} \frac {\sinh \lambda^\prime_i t/2}{\lambda^\prime_i t/2} \;.
\eea

The result for the path-integral (\ref{pathtensor}) is then:
\bea
Z_M \a=\a \int \! dx_0^\mu \int \! d\psi_0^{\ubar \mu} \, (\pi t)^{-\frac d2}
\prod_{i=1}^{d/2} \frac {\lambda_i t/2}{\tanh \lambda_i t/2}
\prod_{i=d/2}^{D/2} \frac {\tanh \lambda^\prime_i t/2}{\lambda^\prime_i t/2}
\prod_{i=d/2}^{D/2} \lambda^\prime_i t/2 \nn \\
\a=\a \int \! dx_0^\mu \int \! d\psi_0^{\ubar \mu} \,
\prod_{i=1}^{d/2} \frac {\lambda_i/2\pi}{\tanh \lambda_i/2\pi}
\prod_{i=d/2}^{D/2} \frac {\tanh \lambda^\prime_i/2\pi}{\lambda^\prime_i/2\pi}
\prod_{i=d/2}^{D/2} \lambda^\prime_i/2\pi
\label{pathten}
\eea
where the last step is valid for the d-form component. This result can be finally rewritten 
as
\be
Z_M = \int_M \frac {\widehat L (R)}{\widehat L (R^\prime)} \wedge e(R^\prime) \;.
\label{Aten}
\ee
Again the realization of $\delta_{\eta}$ does not present particular problems. 
The form of the gauge transformations can be obtained by thinking of the
self-dual tensor as a bispinor \cite{agw}. The operators $\delta_\eta$ and
$\delta_{\eta^\prime}$ are in this case represented by
$$
\delta_\eta = - \eta_\mu \dot x^\mu - D_{\ubar \mu} \eta^\prime_{\ubar \nu} 
\psi_2^{\ubar \mu} \psi_2^{\ubar \nu} \;,\;\; 
\delta_{\eta^\prime} = - \eta^\prime_i \dot x^i
- D_{\ubar i} \eta^\prime_{\ubar j} \psi_2^{\ubar i} \psi_2^{\ubar j}
$$
upon canonical quantization. 
As before, one can exponentiate these operators and expand in normal coordinates. 
The net effect in (\ref{tensoreff}) is again to shift $R_{\mu\nu}
\rightarrow R_{\ubar \mu \ubar \nu} + D_{\ubar \mu} \eta_{\ubar \nu} 
- D_{\ubar \nu} \eta_{\ubar \mu}$ and $R_{\ubar i \ubar j} \rightarrow 
R_{\ubar i \ubar j} + D_{\ubar i} \eta_{\ubar j} - D_{\ubar j}
\eta_{\ubar i}$. Therefore, taking the terms linear in $\eta,\eta^\prime$
again corresponds to take the Wess-Zumino descent, and the final result for the real self-dual
antisymmetric tensor anomaly is
\be
A_M = 2 \pi i \int_M \left[- \frac 18 \frac {\widehat L (R)}{\widehat L (R^\prime)} 
\wedge e(R^\prime) \right]^{(1)} \;.
\label{Aten(1)}
\ee
The additional factor of $1/2$ arises as in \cite{agw}: for the anomaly the relevant component 
of the integrand is the (d+2)-form, whereas it was the d-form component in (\ref{Aten}), so
the rescaling in the second line of (\ref{pathten}) produces an extra factor of $1/2$.

\subsection{Index and G-Index theorems}

In last subsection we have computed the anomalies associated to chiral spinors and 
tensors propagating on a submanifold $M$ of $X$, in interaction with 
gravitational fields propagating on $X$, by working out suitable path integrals.
The results are indices encoding topological data of $M$ and/or $X$.
In order to check the path integral computation and to have a better
understanding of the mathematical nature of the anomalies we have found, we will also
compute directly the indices (\ref{Aspi}),(\ref{Aten}), using the
Atiyah-Singer index theorem and its G-index generalization \cite{AS,ASg}
(see also \cite{ag,fw} for related applications of index theorems).

\subsection{Chiral spinor}

It is well known that the index of the Dirac operator in an even dimensional
manifold $X$ gives the chiral anomaly of a Dirac spinor on $X$. Through the descent
procedure, this is also related to gauge and gravitational anomalies of a chiral spinor in two 
lower dimensions. The close relation between index theory and anomalies can be also 
used in the case we are interested in, i.e. a chiral spinor propagating on a even
dimensional submanifold $M$ of $X$. Although the chiral anomaly (\ref{Aspi})
has been already obtained in \cite{cy} using the family index theorem \cite{ASf},
for completeness we re-derive here that result using the standard index theorem \cite{AS}.

Given the tangent bundle decomposition of $X$ as the Whitney sum of the tangent
and normal bundles of $M$, the corresponding positive and negative chirality spin bundles 
$S^{\pm}_{T(X)}$ decompose as follows in terms of the positive and negative chirality spin 
bundles $S^\pm_{T(M)}$ and $S^\pm_{N(M)}$ lifted from the tangent and normal bundles to $M$:
\be
S^\pm_{T(X)} \rightarrow [\,S^\pm_{T(M)} \otimes S^+_{N(M)} \,] \oplus
[\,S^\mp_{T(M)} \otimes S^-_{N(M)} \,]  \;.
\label{Sbundles}
\ee

The Dirac operator for the charged and reduced fermion we are considering acts on sections
of the bundles (\ref{Sbundles}) tensored with the gauge bundle $V_{\rho}$ in the representation 
$\rho$ in which the fermion transforms, interchanging positive and negative chiralities. More 
precisely, we have the two-term complex
\be
i \Dslash \;:\; \Gamma (M, E^+) \rightarrow \Gamma (M, E^-)
\label{Scomplex}
\ee
where
\be
E^\pm = \left([\, S^\pm_{T(M)} \otimes S^+_{N(M)} \,] \oplus 
[\,S^\mp_{T(M)} \otimes S^-_{N(M)} \,] \right) \otimes V_{\rho} \;.
\label{Scomp}
\ee

It is now straightforward to apply the usual index theorem to the particular case of the two-term 
complex (\ref{Scomplex}). One finds
$$
\mbox{index}(i \Dslash) = (-1)^{\frac {d(d+1)}2} \int_M \mbox{ch}_\rho(V)
\frac {\mbox{ch}(S^+_{T(M)} - S^-_{T(M)}) \,
\mbox{ch}(S^+_{N(M)} - S^-_{N(M)})}{e(T(M))} \mbox{Td} 
(T(M^{C}))
$$
where $\mbox{Td}(T(M^C))$ is the Todd class of the complexified tangent bundle of $M$
and $d={\rm dim}\, M$. The Chern characters of the spin bundles are:
\bea
\a\a \mbox{ch}(S^+_{T(M)} - S^-_{T(M)}) = \prod_{i=1}^{d/2} \, (e^{x_i/2}-e^{-x_i/2}) \;, \nn \\
\a\a \mbox{ch}(S^+_{N(M)} - S^-_{N(M)}) = \prod_{j=d/2}^{D/2} \, \nn
(e^{x_i^\prime/2}-e^{-x_i^\prime/2}) \;,
\eea
with $D={\rm dim}\, X$ and $x_i,x_i^\prime$ respectively the eigenvalues of the
curvature two-form on $T(M)$ and $N(M)$ (note that $x=\lambda/2\pi$ defined before).
By using the standard expressions for the Euler and 
Todd classes, one then easily reproduces (\ref{Aspi}):
\bea
\mbox{index}(i \Dslash) \a=\a \int_M  \, \mbox{ch}_\rho(V) \wedge 
\frac{\widehat{A}(R)}{\widehat{A}(R^{\prime})} \wedge e(R^{\prime}) \;.
\eea

\subsection{Self-dual tensor}

Let us now turn our attention to self-dual tensors. Again, through the descent procedure,
the gravitational anomaly on a 4n+2 dimensional manifold
is related to the index of a classical complex over a 4n+4 dimensional manifold $X$, 
the signature complex:
\be
{\cal D}_+ \;:\;  {}^+\!\!\wedge T^*X \longrightarrow  {}^-\!\!\wedge T^*X \;.
\ee
${\cal D}_+$ maps self-dual forms to anti self-dual forms on $X$.
In the case we are interested, the tensors propagate on $M\subset X$, 
but they are sections of $\wedge\, T^*X$. From the results of last section, 
we learned that a suitable operator that projects onto states propagating
on $M$ only is the ${\bf Z}_2$ operator acting on the normal coordinates:
if $(x^{\mu},y^i)$ are respectively local coordinates on $M$ and its transverse
space in $X$, then
\be
{\bf Z}_2 \;:\; (x^{\mu},y^i) \longrightarrow (x^{\mu},-y^i) \;.
\label{z2}
\ee
What we have to compute is then the so-called $G$-index \cite{ASg} of the signature complex, 
or simply the $G$-signature, where $G={\bf Z}_2$. 
Generically, $G$ can be a compact Lie group acting on $X$ by orientation-preserving 
transformations\footnote{Note that in our cases the transverse space is always
even-dimensional, and (\ref{z2}) is orientation-preserving.} 
(see \cite{sha} for a nice introduction and more
details on the $G$-index). The action of $G$ on $X$ can be also extended 
to vector bundles over $X$, provided that $G$ acts on the bundle $E$
mapping linearly the fiber on the point $x$ to the fiber on $gx$, $\forall g\in G$.
In this case $E$ is also called a $G$-bundle.
Let then $X_G\subset X$ be the subspace left invariant 
by $G$, that is $X_G=\{x\in X: gx=x, \forall g\in G\}$. Let us also denote
with $T_G$ and $N_G$ respectively the tangent and normal bundles of $X_G$ in $X$.
The $G$-signature is then given by (see e.g. \cite{sha})
\be
\mbox{index}({\cal D}_+^G)=\int_{X^G}\, \frac{{\rm ch} (T_G^+-T_G^-)\,
{\rm ch}_G(N_G^+-N_G^-)}{{\rm ch}_G(\tilde{N}_G)\,e(T_G)} {\rm Td}(T_G^{C})
\ee
where $e(T_G)$ and ${\rm Td}(T_G^{C})$ are the usual Euler and Todd classes,
$T_G^{\pm}={}^{\pm}\!\!\wedge T^*X_G$, $N_G^{\pm}={}^{\pm}\!\!\wedge N^*X_G$ and
$\tilde{N}_G=\oplus_i (-)^i \wedge^i N^*X_G$. If $E_G$ is a G-bundle, in general
$E_G=\oplus_i E_G^{(i)}$, where $G$ acts with the element 
$g_i$ in $E_G^{(i)}$. In this case, the Chern character ${\rm ch}_G$ reads
${\rm ch}_G(E_G)=\sum_i {\rm Tr}\,g_i \, \exp \{iF_i/2\pi\}$ 
where $F_i$ is the curvature 2-form on $E_G^{(i)}$.
In our particular case, $G={\bf Z}_2$ and clearly $X_{{\bf Z}_2}=M$,
$T_{{\bf Z}_2}=T(M)$ and $N_{{\bf Z}_2}=N(M)$.
$T(M)$ and $N(M)$ are $G$-bundles in which ${\bf Z}_2$ acts
respectively with the elements I and -I. One then finds
\bea
\a\a{\rm ch} (T_G^+-T_G^-) = \prod_{i=1}^{d/2} \, (e^{x_i}-e^{-x_i}) \;,\;\;
{\rm ch}_{G}(N_G^+-N_G^-) = \prod_{j=d/2}^{D/2} \, (e^{x_i^{\prime}}-e^{-x_i^{\prime}}) \;, \nn \\
\a\a{\rm ch}_G(\tilde{N}_G) = \prod_{j=d/2}^{D/2} \, (1+e^{x_i^{\prime}})\,(1+e^{-x_i^{\prime}}) \;, \nn
\eea
where the eigenvalues $x_i,x_i^{\prime}$ are defined as previously. Putting all together, one 
finally reproduces eq.(\ref{Aten}):
\be
\mbox{index}({\cal D}_+^{{\bf Z}_2}) =
\int_M \frac{\widehat{L}(R)}{\widehat{L}(R^{\prime})} \wedge e(R^{\prime}) \;.
\ee


\section{Inflow on D-brane and O-plane intersections}
\setcounter{equation}{0}

In previous sections, we have shown that the anomaly polynomials for the chiral spinors and 
self-dual tensors living on overlapping D-brane/O-planes are given by
\bea
I_{1/2}^\rho (F,R,R^\prime) \a=\a \mbox{ch}_\rho (F) \wedge 
\frac {\widehat A (R)}{\widehat A (R^\prime)} \wedge e(R^\prime) \;,
\label{Ispinor}\\
I_{A}(R,R^\prime) \a=\a - \frac 18 \frac {\widehat L (R)}{\widehat L (R^\prime)} \wedge e(R^\prime) \;.
\label{Itensor}
\eea
In this section we show that these anomalies are exactly cancelled by the inflow of anomaly
associated to the anomalous couplings (\ref{WZD}) and (\ref{WZO}). To this aim, we shall 
briefly recall how the inflow mechanism works in the general case, following \cite{cy}.

Consider a set of defects $M_i$ in spacetime $X$ with anomalous couplings of the form
\be
\label{ano}
S = - \sum_i \frac {\mu_i}2 \int_{M_i} C \wedge Y_i \;.
\ee
By integrating by parts, the integrand can be rewritten in terms of the constant parts 
$Y_{i0}$, which we set to 1 by suitably normalizing the charges $\mu_i$, and the 
descents $Y_i^{(0)}$, as $C \mp H \wedge Y_i^{(0)}$, the sign depending on whether 
$C$ contains even (Type IIB) or odd (Type IIA) forms.
The complete action for the RR fields in presence of this sources can then be written as 
an integral over all of spacetime $X$ by using a current representative $\tau_{M_i}$
in the space which is dual in $X$ to the forms on $M_i$:
\be
S= - \frac 14 \int_X H \wedge {}^*H - \sum_i \frac{\mu_i}2 \int_X \tau_{M_i}
\left(C \mp H \wedge Y_i^{(0)} \right) \;.
\label{actionRR}
\ee
$\tau_{M_i}$ is itself a form of rank equal to the codimension $D-d_i$ of $M_i$ in $X$. 
Locally, it can be represented by a generalization to forms of Dirac's $\delta$-function given
by $\tau_{M_i} \sim \delta(x^{d_i}) ... \delta(x^D) dx^{d_i} \wedge ... \wedge dx^D$. 
Globally, it is however a section of the normal bundle $N(M_i)$.
The equations of motion and Bianchi identity implied by (\ref{actionRR}) are
\bea
d^*H \a=\a \sum_i \mu_i \tau_{M_i} \wedge Y_i \;,
\label{motion} \\
d H \a=\a - \sum_i \mu_i \tau_{M_i} \wedge \bar  Y_i \;,
\label{bianchi}
\eea
where $\bar  Y_i$ is obtained from $Y_i$ by complex conjugation of the gauge group 
representation.
Moreover, for consistency one must have vanishing total charge for the top RR form.
It is clear form the modified Bianchi identity (\ref{bianchi}) 
that the field-strength $H$ cannot be identified anymore with $dC$. 
Rather, the minimal solution of  (\ref{bianchi}) is
$H= dC \mp \sum_i \mu_i \tau_{M_i} \wedge \bar  Y_i^{(0)}$.
Since $H$, being a physical observable, 
must be gauge invariant, $C$ must acquire an anomalous gauge transformation
to compensate the gauge variation of the second term:
$\delta_\eta C = \sum_i \mu_i \tau_{M_i} \wedge \bar  Y_i^{(1)}$.
Consequently, under a gauge transformation $\delta_\eta$ the couplings (\ref{ano}) present
an anomaly given by
\be
\delta_\eta (iS) = - \,i\sum_{i,j} \frac {\mu_i \mu_j}2 \int_X \tau_{M_i} \wedge \tau_{M_j} 
\left(Y_i \wedge \bar  Y_j \right)^{(1)} \;.
\ee
All the anomaly is localized on the intersections $M_{ij}$ of pairs of 
defects $M_i$ and $M_j$. In order to see this explicitly and correctly, 
remember that $\tau_{M_i}$ are sections of the normal bundles $N(M_i)$,
with compact support on it. The current is then well defined on $N(M_i)$
and represented on $M_i$ by taking the zero section of $N(M_i)$.
It is now a standard result (see for instance \cite{bott}) that in cohomology
$\tau_{M_i}$ can be identified with the Thom class $\Phi[N(M_i)]$ of $N(M_i)$,
whose zero section is the Euler class of $N(M_i)$.
This implies the following property for the currents \cite{cy}:
$\tau_{M_i} \wedge \tau_{M_j} = \tau_{M_{ij}} \wedge e(N(M_{ij}))$.
Using the freedom left over in the descent procedure, the inflow can then be rewritten as
\be
\delta_\eta (iS) = - \,i\sum_{i,j} \frac {\mu_i \mu_j}2 \int_{M_{ij}}
\left[\left(Y_i \wedge \bar Y_j \right) \wedge e(N_{ij}) \right]^{(1)} \;.
\ee

It is now straightforward to use this result to show that the anomalies (\ref{Ispinor}) and 
(\ref{Itensor}) are cancelled by the inflows on BB, BO and OO intersections. Specializing 
to two overlapping defects on the same manifold $M$, the inflow of anomaly can be written 
as $A_{ij}=2 \pi i \int_{M_{ij}} I_{ij}^{(1)}$ with
\be
I_{ij} = - \frac {\mu_i \mu_j}{4 \pi} Y_i  \wedge \bar Y_j \wedge e(N_{ij}) \;.
\label{ij}
\ee

We set $d=p+1$ but keep $D$ generic, and use the couplings (\ref{WZD}) and 
(\ref{WZO}) for a Dp-brane and an Op-plane. It is easily seen that the $d$-form part in 
(\ref{ij}) has precisely the right powers of $(4 \pi^2 \al)$ to cancel the  factor 
$(4 \pi^2 \al)^{-(p+1)/2}$ in the charges $\mu_{i,j}$. The effective Dp-brane and Op-plane
charges are then $\mu_p =\alpha  \sqrt{2 \pi}$ and 
$\mu_p^\prime = - 2^{p+1 - D/2} \alpha \sqrt{2\pi}$.
The numerical coefficient $\alpha$ depends on the particular model, and we will derive it
case by case. The anomaly inflows on the BB, BO and OO are the following:

\vskip 9pt 
\noindent 
{\bf BB intersection}
\vskip 1pt 
\noindent
Using the property $\mbox{ch}_{\rho_1} (F) \wedge \mbox{ch}_{\rho_2} (F) = 
\mbox{ch}_{\rho_1 \otimes \rho_2} (F)$, one finds
\be
I_{BB} (F,R,R^\prime) = - \frac {\alpha^2}2 \mbox{ch}_{\lambda \otimes \bar \lambda} (F) \wedge
\frac {\widehat A (R)}{\widehat A (R^\prime)} \wedge e(R^\prime)
\ee
that is
\be
I_{BB} (F,R,R^\prime) =- \frac {\alpha^2}2 I_{1/2}^{\lambda \otimes \bar \lambda} (F,R,R^\prime) \;.
\ee

\vskip 9pt 
\noindent 
{\bf BO intersection}
\vskip 1pt 
\noindent
Using the relations $\mbox{ch}_{\rho_1} (F) + \mbox{ch}_{\rho_2} (F) = 
\mbox{ch}_{\rho_1 \oplus \rho_2} (F)$ and $\widehat A(R) \wedge \widehat L(R/4) = 
\widehat A(R/2)$, one finds in this case
\bea
I_{BO+OB} (F,R,R^\prime) \a=\a 2^{p +1- \frac D2} \frac {\alpha^2}2 
\mbox{ch}_{\lambda \oplus \bar \lambda} (F) 
\wedge \frac {\widehat A (R/2)}{\widehat A (R^\prime/2)} \wedge e(R^\prime) \nn \\
\a=\a \frac {\alpha^2}4 \mbox{ch}_{\lambda \oplus \bar \lambda} (2 F) 
\wedge \frac {\widehat A (R)}{\widehat A (R^\prime)} \wedge e(R^\prime) \;.
\eea
The second line is obtained by rescaling the argument of the Euler class by $1/2$,
producing a factor $2^{\frac {D-p-1}2}$, and then rescale all the arguments by factor $2$, 
giving an additional factor $2^{-\frac {p+3}2}$ for the relevant (p+3)-form component. 
Therefore
\be
I_{BO+OB} (F,R,R^\prime) = \frac {\alpha^2}4 I_{1/2}^{\lambda \oplus \bar \lambda} 
(2F,R,R^\prime) \;.
\ee

\vskip 9pt 
\noindent 
{\bf OO intersection}
\vskip 1pt 
\noindent
One finds in this case
\bea
I_{OO} (F,R,R^\prime) \a=\a - 4^{p+1 - \frac D2} \frac {\alpha^2}2
\frac {\widehat L (R/4)}{\widehat L (R^\prime/4)} \wedge e(R^\prime) \nn \\
\a=\a - \frac {\alpha^2}8 \frac {\widehat L (R)}{\widehat L (R^\prime)} \wedge e(R^\prime) 
\eea
where the second equality follows from manipulations similar to those performed before. 
Then
\be
I_{OO} (R,R^\prime) = \alpha^2 I_{A} (R,R^\prime) \;.
\ee

\vskip 10pt

This demonstrate that the anomaly inflow on D-branes and O-planes intersections have the 
required form to cancel the anomaly of the fields living on them. 
One has chiral spinors in the representation $\lambda \otimes \bar \lambda$ of the gauge group 
for BB, essentially a chiral spinor in the representation $\lambda \oplus\bar \lambda$ 
of the gauge group for BO, and a self-dual antisymmetric tensor, neutral under the 
gauge group, for OO. The precise coefficients depend on the particular model through the 
parameter $\alpha$. We will show that they are indeed correct and discuss them in more detail for 
some particular Type IIB orientifolds in next section. 


\section{Anomaly cancellation in Type IIB orientifolds}
\setcounter{equation}{0}

In this section, we shall discuss some simple examples of Type IIB orientifold models and 
discuss anomaly cancellation in the light of our results. 
In the following we will consider two simple examples: type I theory in $D=10$ and the 
$T^4/{\bf Z_2}$ orientifold model in $D=6$ dimensions \cite{PS,GP}. Although in both cases the 
cancellation of spacetime anomalies is well understood \cite{GS, americani}, our aim will 
be simply to reinterpret those results as special cases of the inflow mechanism discussed 
in the previous sections. In particular, we will explicitly show that, as mentioned in the introduction,
the anomaly coming from neutral states combines into that of a self-dual antisymmetric 
tensor (OO), whereas the one from charged states can be recast into that of chiral spinors, 
essentially in the bifundamental (BB) and fundamental (BO) representations of the gauge group.
Moreover, the coefficients will turn out to be precisely those required to cancel these anomalies 
by the inflow mechanism. We will also briefly discuss the anomaly coming from neutral fields 
in other $T^4{\bf Z}_N$ orientifolds. A detailed and complete account of how this mechanism can 
be extended to these and more general six-dimensional models \cite{GJ,DP,ABPSS}, for which an 
analysis of anomaly cancellation is still lacking, will be reported elsewhere.

In the following, we will make use of the following relations between the anomaly polynomial of 
different fields:
\bea
\a\a D=10: \hspace{5pt} I_{1/2}(R) - I_{3/2}(R) - I_{A}(R) = 0 \label{D10} \\
\a\a D=6: \hspace{10pt} 21 \, I_{1/2}(R) - I_{3/2}(R) + 8 \, I_{A}(R) = 0 \raisebox{17pt}{} 
\label{D6}
\eea
where $I_{1/2}(R)$, $I_A(R)$ and $I_{3/2}(R)$ are the standard gravitational anomaly polynomials for 
spinors, antisymmetric tensors and Rarita-Schwinger fields \cite{agw}.

\subsection{Type I theory}

Consider first Type I theory in $D=10$ as the simplest Type IIB orientifold \cite{Sag}. Taking 
$n_9$ D9-branes together with 1 O9-plane, the gauge group would be $SO(n_9)$, and consistency of 
the theory requires $n_9=32$ \cite{GS}. 

\vskip 9pt
\noindent 
{\bf One-loop anomalies}
\vskip 1pt 
\noindent
Keeping $n_9$ arbitrary, the anomalous fields in the 
model are the following
$$
\begin{array}{ll}
\mbox{\hspace{-215pt}$\bullet$ 1 grav. mult.:}&
\displaystyle{-\frac 12I_{1/2}(R) + \frac 12I_{3/2}(R)} \\
\mbox{\hspace{-215pt}$\bullet$ 1 vec. mult.:}&
\displaystyle{\frac 12I_{1/2}^{{\bf \frac {n_9(n_9-1)}2}}(F,R)}
\raisebox{20pt}{}
\end{array}
$$
where the factors $1/2$ are due to the fact that all fermions are real.
Using (\ref{D10}), the total anomaly from neutral fields is therefore
\be
I_{n}(R) = - \frac 12 I_{A}(R) \;.
\label{n10}
\ee
For the charged fields, it is convenient to rewrite the trace in the adjoint
${\bf n_9(n_9-1)/2}$ in terms of traces in the fundamental ${\bf n_9}$. It is not difficult to verify
order by order that for $SO(n)$ one has
\be
\mbox{ch}_{\bf \frac {n(n-1)}2} (F) =\frac 12 \left(\mbox{ch}_{\bf n \otimes n} (F) - 
\mbox{ch}_{\bf n} (2 F) \right) \;.
\ee
The anomaly from charged fields can then be written as
\be
I_{c}(F,R) = \frac 14 I_{1/2}^{\bf n_9 \otimes n_9}(F,R) - \frac 14 I_{1/2}^{\bf n_9}(2 F,R) \;.
\label{c10}
\ee

\vskip 15pt 
\noindent 
{\bf Anomaly inflow}
\vskip 1pt 
\noindent
Consider now the inflows on the D9D9, D9O9 and O9O9 intersections. The Chan-Paton 
representation $\lambda$ is the fundamental $\bf n_9$. Using then the formulae 
derived in last section one finds
\bea
\a\a I_{BB} (F,R) = -\frac {\alpha^2}2 I_{1/2}^{\bf n_9 \otimes n_9} (F,R) \;, \nn \\
\a\a I_{BO+OB} (F,R) = \frac {\alpha^2}2 I_{1/2}^{\bf n_9} (2F,R) \;, \nn \raisebox{15pt}{} \\
\a\a I_{OO} (R) = {\alpha^2} I_{A} (R) \raisebox{15pt}{} \;.
\eea
We see therefore that the inflow cancels precisely the anomalies (\ref{n10}) and (\ref{c10}) 
only if $\alpha = 1/\sqrt{2}$.
This is indeed the correct value for Type I D-branes and O-planes \cite{bach}.

We have therefore verified in this simple case that the inflow on BB, BO and OO exactly 
cancels respectively the anomalies of the charged and neutral fields. Moreover,
it is very clear from this way of doing that the requirement $n_9 =32$ appears exclusively
from charge cancellation. Notice also that the requirement of vanishing irreducible
terms in the anomaly polynomial ${\rm tr} F^6, {\rm tr}R^6$ is equivalent to charge cancellation 
for the 10-form RR potential. Indeed, the inflow necessary to cancel these terms would involve 
the 10-form and require the presence of a clearly inexistent magnetic dual (-2)-form 
with anomalous couplings proportional to $C_{(-2)} \wedge {\rm tr}F^6$, 
$C_{(-2)} \wedge {\rm tr}R^6$ \cite{polcai}. 

\subsection{K3 orientifolds}

\vskip 2pt

Consider now $T^4/{\bf Z}_N$ orientifolds of Type IIB theory, which can be interpreted as
generalizations of $K3$ compactifications of Type I theory to $D=6$.
The low energy effective theory 
has $N=1$ $D=6$ supersymmetry, and beside the usual gravitational and tensor multiplets
of $N=1$ $D=6$ supergravity, it will involve a vector multiplet in the adjoint representation of
the gauge group and a certain number of charged hyper and neutral tensor matter multiplets,
depending on the model. Indicating with $n_H$ the number of hyper multiplets and $n_T$ the
number of additional tensor multiplets, the total anomaly from neutral fields is found to be,
using (\ref{D6}),
\be
I_n (R) = (n_T - 8) I_A (R) + (n_T + n_H - 20) I_{1/2}(R) \;.
\ee
As pointed out in \cite{GJ} all consistent ${\bf Z}_N$ orientifold models have $n_T + n_H = 20$. 
This condition is closely related to the geometric properties of the underlying $K3$ surface. 
Additional models constructed as open descents of Gepner models \cite{ABPSS} 
also satisfy this condition. It is remarkable that the total neutral field anomaly has the form of 
the anomaly of a self-dual tensor:
\be
I_n (R) = (n_T - 8) I_A (R) \;.
\label{Inzn}
\ee

The total anomaly coming from charged fields must be analyzed model by model.
The simplest example we shall consider in the following is the ${\bf Z}_2$ orientifold with 
maximally enhanced gauge group \cite{GP}. Taking $2 n_9$ D9-branes and $2 n_5$ D5-branes, 
together with 1 O9-plane and 16 O5-planes, 
with all the $2 n_5$ D5-branes at a single fixed-point on top of the corresponding O5-plane, 
the gauge group is $U(n_5) \times U(n_9)$, and consistency of the theory requires 
$n_5=n_9=16$ \cite{GP}. Anomaly cancellation in this model has been studied in \cite{americani},
where it was shown that the anomaly factorizes and is cancelled by a generalization of the 
GS mechanism. We will give here an interpretation in terms of inflows on D-brane/O-plane
intersections.

\vskip 9pt 
\noindent 
{\bf One-loop anomalies}
\vskip 1pt 
\noindent
Keeping $n_{5,9}$ arbitrary, the anomalous fields in the ${\bf Z}_2$ model are the following
$$
\begin{array}{ll}
\mbox{\hspace{-0pt}$\bullet$ 1 grav. mult.:}& -I_{3/2}(R) - I_A (R) \\
\mbox{\hspace{-0pt}$\bullet$ 1 tens. mult.:}& I_A (R) + I_{1/2} (R) \raisebox{20pt}{} \\
\mbox{\hspace{-0pt}$\bullet$ 1 vec. mult. in the $({\bf n_5^2}, {\bf 1}) 
\oplus ({\bf 1},{\bf n_9^2})$:}& 
-I_{1/2}^{({\bf n_5^2},{\bf 1}) \oplus ({\bf 1},{\bf n_9^2})}(F,R) \raisebox{20pt}{} \\
\mbox{\hspace{-0pt}$\bullet$ 2 hyp. mult. in the 
$({\bf \frac {n_5(n_5-1)}2}, {\bf 1}) \oplus  ({\bf 1},{\bf \frac {n_9(n_9-1)}2})$:}&
2 \, I_{1/2}^{({\bf \frac {n_5(n_5-1)}2}, {\bf 1}) \oplus ({\bf 1},{\bf \frac {n_9(n_9-1)}2})}(F,R) 
\raisebox{20pt}{} \\
\mbox{\hspace{-0pt}$\bullet$ 1 hyp. mult. in the $({\bf n_5}, {\bf n_9})$:}& 
I_{1/2}^{({\bf n_5},{\bf n_9})}(F,R) \raisebox{20pt}{} \\
\mbox{\hspace{-0pt}$\bullet$ 20 hyp. mult. in the $({\bf 1}, {\bf 1})$:}& 20 \, I_{1/2} (R) 
\raisebox{20pt}{}
\end{array}
$$
Using (\ref{D6}), the total anomaly from neutral fields is therefore
\be
I_{n}(R) = - 8 I_{A}(R)
\label{In}
\ee
which is a particular case of (\ref{Inzn}) with $n_T=0$.
For the charged fields, it is as usual convenient to rewrite the traces in all the representations 
in terms of traces in the fundamental representations. For $U(n)$ one has obviously
\be
\mbox{ch}_{\bf n^2} (F) = \mbox{ch}_{{\bf n} \otimes {\bf \bar n}} (F)
\ee
for the adjoint representation, and one can check order by order that
\be
\mbox{ch}_{\bf \frac {n(n\pm1)}2} (F) = \frac 12 \left(\mbox{ch}_{{\bf n} \otimes {\bf n}} (F) \pm 
\mbox{ch}_{\bf n} (2 F) \right)
\ee
for the symmetric and antisymmetric tensor representations.

Using the fact that only even powers of the field strength $F$ appear in the anomaly polynomial, 
the total anomaly from charged fields is then found to be (the symbol $\ominus$ means that one 
has to take the differences of the Chern classes in the two representations):
\bea
I_{c}(F,R) \a=\a \frac 14 I_{1/2}^{({\bf n_5} \oplus {\bf \bar n_5},{\bf n_9} \oplus {\bf \bar n_9})}
(F,R) - \frac 12 \left(I_{1/2}^{({\bf n_5} \oplus {\bf \bar n_5},{\bf 1})} +
I_{1/2}^{({\bf 1},{\bf n_9} \oplus {\bf \bar n_9})} \right) (2F,R) \label{Ic} \\
\a\;\a +  \frac 14 \left(I_{1/2}^{({\bf n_5} \ominus {\bf \bar n_5}, {\bf n_9} \ominus {\bf \bar n_9})}
+2  I_{1/2}^{(({\bf n_5} \ominus {\bf \bar n_5}) \otimes ({\bf n_5} \ominus {\bf \bar n_5}),{\bf 1})} 
+2  I_{1/2}^{({\bf 1}, ({\bf n_9} \ominus {\bf \bar n_9}) \otimes ({\bf n_9} \ominus {\bf \bar n_9}))} 
\right) (F,R) \;. \nn
\eea
The first three terms contain only $\mbox{tr} F^{2m}$ factors, whereas the last three
contain only $\mbox{tr} F^{2m+1}$ factors. The latter are therefore entirely responsible 
for pure Abelian and mixed Abelian-non Abelian anomalies.

\vskip 9pt 
\noindent 
{\bf Anomaly inflow}
\vskip 1pt 
\noindent
Consider now the inflows on the DpDq, DpOq and OpOq intersections, with $p,q=5,9$. 
Due to the ${\bf Z}_2$ projection and the consequent appearance of twisted closed strings,
there will be in this case two types of anomaly inflow associated to magnetic interactions 
of the various D-branes and O-planes. The first one is the usual one, involving the 
exchange of untwisted RR forms and the anomalous couplings (\ref{WZD}) and (\ref{WZO})
to them. The second one involves the exchange of twisted RR-forms and additional anomalous 
couplings to them, as shown in \cite{americani}. 

Consider first the inflow from the untwisted sector. The Chan-Paton representation $\lambda$ is 
in this case $({\bf n_5} \oplus {\bf \bar n_5},{\bf 1}) \oplus ({\bf 1},{\bf n_9} \oplus {\bf \bar n_9})$. 
Using then the formulae derived in last section, one can then compute the 
inflow directly in terms of D=10 RR fields and then integrate over the compact space. Since we have 
already explicitly taken into account the singularities at the 16 orbifold fixed-points through 
16 O5-planes in the background, we have to integrate over the flat torus $T^4$, rather than 
the orbifold $T^4/{\bf Z}_2$; otherwise, one would overcount the singularities\footnote{This can be 
also seen by integrating the anomalous coupling (\ref{WZO}) for the O9-plane of Type I theory
on $T^4/{\bf Z}_2$, where the anomalous couplings for the 16 O5-planes in the ${\bf Z}_2$ model 
appear in the fixed-point contributions to $\int_{T^4/{\bf Z}_2} \sqrt{\widehat{L}}$.}. 
The D9D9, D9O9 and O9O9 inflows vanish when integrated over $T^4$. The D5D5, D5O5 and O5O5 
inflows also vanish, since the normal bundle to them is trivial and the corresponding anomalous 
couplings vanish as well. The only non-vanishing inflows come therefore from the D5D9, D5O9, O5D9 
and O5O9 intersections. The normal bundle is null, and all the corresponding characteristic 
classes are equal to 1. One then finds in total
\bea
\a\a I_{D5D9}^{(un.)} (F,R) = - \alpha^2 I_{1/2}^{({\bf n_5} \oplus {\bf \bar n_5},
{\bf n_9} \oplus {\bf \bar n_9})} (F,R) \;, \nn \\
\a\a I_{D5O9+D9O5}^{(un.)} (F,R) = 2 \alpha^2 \left(
I_{1/2}^{({\bf n_5} \oplus {\bf \bar n_5},{\bf 1})}+ 
I_{1/2}^{({\bf 1},{\bf n_9} \oplus {\bf \bar n_9})} \right) (2F,R) \;, \nn \raisebox{15pt}{} \\
\a\a I_{O5O9}^{(un.)} (R) = 32 \alpha^2 I_{A} (R) \;. \raisebox{15pt}{}
\label{inflowun}
\eea

Consider now the inflow from the twisted sector. It was shown in \cite{americani} that this 
cancels the Abelian anomaly coming from the charged hyper multiplets in the spectrum. 
The corresponding anomalous couplings are also responsible for a spontaneous breaking 
of $U(n_5) \times U(n_9)$ to $SU(n_5) \times SU(n_9)$. The gauge field dependence of these 
couplings was inferred in \cite{americani}. The complete result can be obtained by factorizing 
twisted sector magnetic interactions in the odd spin-structure,
along the lines of \cite{mss}. The M\"obius strip and Klein bottle amplitudes giving the BO
and OO twisted magnetic interaction vanish trivially. One can thus immediatly conclude 
that O-planes do not have anomalous couplings to twisted RR forms. For the cylinder
encoding BB twisted magnetic interactions, the gravitational part is unaltered, whereas the 
net effect of the twist is a conjugation of Chan-Paton wave-function through the symplectic matrix
\be
M = \pmatrix{ 0 \a I_{n_{5}} \a 0 \a 0 \cr - I_{n_5} \a 0 \a 0 \a 0 \cr
0 \a 0\a 0 \a I_{n_{9}} \cr 0 \a 0 \a - I_{n_9} \a 0 \cr} \;.
\ee
D5-branes at fixed point $I$ and D9-branes have therefore the following anomalous couplings to 
twisted RR forms $C_I^{tw.}$:
\bea
S^{tw.}_{D_{5}} \a=\a \frac {\tilde \mu_{5}}2 \int \! C_I^{tw.} \wedge\, 
\mbox{ch}_{({\bf n_5} \oplus {\bf \bar n_5}, {\bf 1})} (M F)\, 
\wedge \, \left.\sqrt{\widehat{A}(R)}\right|_{6-form} \nn \\
S^{tw.}_{D_{9}} \a=\a \frac {\tilde \mu_{9}}8 \sum_{I=1}^{16} \int \! C_I^{tw.} \wedge\, 
\mbox{ch}_{({\bf 1}, {\bf n_9} \oplus {\bf \bar n_9})} (M F)\, 
\wedge \, \left.\sqrt{\widehat{A}(R)}\right|_{6-form}
\label{WZDtw}
\eea
where $\tilde \mu_{5,9}=\beta \sqrt{2\pi}$ and the integral is over the (5+1)-dimensional 
non-compact space. $\beta$ is again a numerical coefficient, which will be fixed in the following.
A D5-brane at fixed-point I couples therefore only to $C^{tw.}_I$, whereas
a D9-brane wrapped on compact space couples to all the 16 $C^{tw.}_I$'s \cite{americani}.
The net effect of the matrix $M$ in the Chern character is:
\be
\mbox{ch}_{({\bf n_5} \oplus {\bf \bar n_5},{\bf 1})} (M F) = 
\mbox{ch}_{({\bf n_5} \ominus {\bf \bar n_5},{\bf 1})} (F) \;,\;\;
\mbox{ch}_{({\bf 1},{\bf n_9} \oplus {\bf \bar n_9})} (M F) = 
\mbox{ch}_{({\bf 1},{\bf n_9} \ominus {\bf \bar n_9})} (F) \;.
\ee
Due to this property, it is evident that only odd powers of $F$ appear in (\ref{WZDtw}). 
Correspondingly, the associated twisted RR forms are 0-forms and their magnetic dual 4-forms,
responsible for the inflow in this sector. The final result for the twisted inflow on BB intersections 
is then
\bea
I_{(D5 + D9)(D5 + D9)}^{(tw.)} (F,R) \a=\a  - \frac {\beta^2}4 \left[
I_{1/2}^{({\bf n_5} \ominus {\bf \bar n_5},{\bf n_9} \ominus {\bf \bar n_9})}(F,R) 
\right.\label{inflowtw} \\ 
\a\,\a \hspace{28pt} \left. +2 \left(I_{1/2}^{(({\bf n_5} \ominus {\bf \bar n_5})  \otimes 
({\bf n_5} \ominus {\bf \bar n_5}),{\bf 1})} + I_{1/2}^{({\bf 1},({\bf n_9} 
\ominus {\bf \bar n_9}) \otimes ({\bf n_9} \ominus {\bf \bar n_9}))}\right) (F,R) \right] \nn
\eea
whereas the twisted inflow on BO and OO intersections vanish. 

We see that the inflows (\ref{inflowun}) from the untwisted sector cancel precisely
the one loop anomaly (\ref{In}) of neutral fields and the non-Abelian one of charged 
fields in eq.(\ref{Ic}), if one takes $\alpha=1/2$. This is indeed the minimal value required
by the D1-D5 Dirac quantization condition, since in this model the D5-branes are grouped into 
sets of 4 to make a dynamical D5-brane, due to the orientifold projection: 
$4 (\alpha \sqrt{2\pi})^2 = 2 \pi$. Similarly, the inflow (\ref{inflowtw}) from the twisted sector
cancels the remaining part of the anomaly in eq.(\ref{Ic}), if one takes $\beta = 1$. 

Summarizing, we have shown that the anomalous couplings required to cancel all the one-loop 
anomalies in the model do indeed arise. Similarly to the Type I case, the condition $n_5=n_9=16$
comes from the requirement of vanishing irreducible terms in the anomaly polynomial,
again because these would require unphysical propagating negative forms.
This confirms what found in \cite{americani}.


\section{Conclusions}
\setcounter{equation}{0}

In this paper we have computed the anomalies for reduced chiral spinors and self-dual 
antisymmetric tensors living on D-brane/O-plane intersections and showed that these
are cancelled through the inflow mechanism induced by the couplings 
(\ref{WZD}) and (\ref{WZO}). 
The main point is that in any consistent string theory model, the one-loop 
anomaly can be recast into a very particular form. The part arising from charged fields
in the open string sector can be recast into the anomaly of a charged spinor in appropriate 
representations of the gauge group, whereas the part coming from neutral fields
in the closed string sector has to combine into that of neutral self-dual antisymmetric 
tensors. The inflows on BB, BO and OO intersections then cancel these anomalies. 
The condition that the irreducible part of the anomaly polynomial cancels is mapped to the 
absence of inflow involving non-existent negative forms.

The relation between inflow and anomalies is very clear in string theory, where the two are
related by the usual open-closed duality in the odd spin-structure of potentially divergent 
annulus, M\"obius strip and Klein bottle diagrams involving D-branes and O-planes. 
In the open string channel, they are interpreted as anomalous one-loop amplitude on the 
worlvolumes of the corresponding D-branes and/or O-planes, whereas in the closed string 
channel they correspond to anomalous magnetic interactions responsible for the inflow
mechanism. Extending \cite{GS}, one could imagine to compute anomalies in string theory, 
by studying one-loop correlation functions with one unphysical external particle probing the 
breakdown of gauge-invariance. Presumably, the only effect of the unphyiscal vertex will be to
implement the descent on the correlation of the physical vertices. Along the lines of 
\cite{mss}, the remaining correlation can be exponentiated, reducing the amplitude to
an effective supersymmetric partition function in the odd spin-structure. At that point,
due to the topological nature of the amplitude, one can reduce the (1+1)-dimensional
$\sigma$-model to 0+1, the computation boiling then down to that of Section 2.1.

It will be very interesting to discuss string theory compactifications in which the anomaly 
associated to the normal bundle is potentially non-vanishing. For instance, this is the case
of D-branes, and eventually O-planes, wrapped on supersymmetric cycles of Calabi-Yau 
manifolds. Even more interestingly, gravity in transverse space seems to induce consistent
gauge-like couplings for antisymmetric tensors. In presence of a non-trivial normal bundle,
this allows, according to eq.(\ref{Aten(1)}), mixed anomalies for self-dual tensors in 
4n dimensions.


\vspace{5mm}
\par \noindent {\large \bf Acknowledgments}
\vspace{3mm}

We would like to thank D. Bernardini, R. Dijkgraaf, E. Scheidegger, S. Theisen and A.K. Waldron
for very interesting suggestions and many useful discussions.
This work has been supported by EEC under TMR contract ERBFMRX-CT96-0045 and 
by the Nederlandse Organisatie voor Wetenschappelijk Onderzoek (NWO). 


\end{document}